# Linear decrease of critical temperature with increasing Zn substitution in the iron-based superconductor BaFe$_{1.89-2x}$Zn$_{2x}$Co$_{0.11}$As$_2$

Jun Li,[1,2,*] Yanfeng Guo,[3] Shoubao Zhang,[3] Shan Yu,[1] Yoshihiro Tsujimoto,[3] Hiroshi Kontani,[5] Kazunari Yamaura,[1,2,4,*] Eiji Takayama-Muromachi [2,3,4]

[1] Superconducting Properties Unit, National Institute for Materials Science, 1-1 Namiki, Tsukuba, Ibaraki 305-0044, Japan

[2] Department of Chemistry, Graduate School of Science, Hokkaido University, Sapporo, Hokkaido 060-0810, Japan

[3] International Center for Materials Nanoarchitectonics (MANA), National Institute for Materials Science, 1-1 Namiki, Tsukuba, Ibaraki 305-0044, Japan

[4] JST, Transformative Research-Project on Iron Pnictides (TRIP), 1-1 Namiki, Tsukuba, Ibaraki 305-0044, Japan

[5] Department of Physics, Nagoya University, Furo-cho, Nagoya 464-8602, Japan

The nonmagnetic impurity effect is studied on the Fe-based BaFe$_{1.89}$Co$_{0.11}$As$_2$ superconductor ($T_c$ = 25 K) with Zn substitution for Fe up to 8 at. %, which is achieved by means of high-pressure and high-temperature heating. $T_c$ decreases almost linearly with increasing the Zn content and disappears at ~8 atomic %, being different in the shared phenomenology of the early Zn doping studies, where $T_c$ decreases little. The $T_c$ decreasing rate, however, remains much lower (3.63 K/%) than what is expected for the $s_\pm$-wave model, implying the model is unlikely. Another symmetry model such as the non-sign reversal $s$-wave model may better account for the result.

Discovery of the Fe-based superconductor in 2008 opened a fundamental question about the pairing symmetry of its superconductivity (SC).[1,2] In the compound, fully-gapped multiband superconductivity is realized according to NMR,[3] angle-resolved photoemission spectroscopy,[4] London penetration depth meserments,[5] and μSR studies.[6] Several independent groups theoretically predicted a sign-reversal s-wave model ($s_\pm$-wave) for the SC,[2] and recent half-flux quantum experiments on $NdFeAsO_{0.88}F_{0.12}$ supported the model.[7] Meantime, a non-sign-reversal s-wave model ($s_{++}$-wave) was proposed to account for the hump structure observed in neutron-scattering measurements below $T_c$.[8] The $s_{++}$-wave model is highly developed and is competitive with the $s_\pm$-wave model.[9-11] Besides, a d-wave model is still competing with the $s_{++}$- and $s_\pm$-wave models.[12]

The $s_{++}$- and $s_\pm$-wave share the same sign for the hole-Fermi pockets, but do not for the electron-Fermi pockets. The d-wave state has opposite signs for the nearest-neighbor electron-Fermi pockets. Because nonmagnetic impurity (NMI) causes pair-breaking in different ways depending on the pairing symmetry, thus the NMI study is expected to greatly help to answer the open question.[13-16] Anderson's theorem predicts that a NMI does not break pairing in an isotropic non-sign-reversal SC state, while it does in an anisotropic state.[16] The theorem indeed well describes the results for such as the cuprate superconductor, which quickly loses the SC by a small amount of NMI.[13]

Since $Zn^{2+}$ has a tightly closed d shell, a doped Zn normally works as a better NMI. A few at.% of Zn in fact acts as a strong scattering center in a superconductor, though it has little influence on the magnetism and transport properties.[13-15] Because the doped Zn actually plays a crucial role of the pairing symmetry determination, we may expect that it works as well in the Fe-based superconductor. In early studies, Cheng et al. reported that the doped Zn hardly affects the SC of the p-type $Ba_{0.5}K_{0.5}Fe_2As_2$,[17] as does Li et al. for the n-type $LaFeAsO_{0.85}F_{0.15}$.[18] However, the SC is completely suppressed by at most 3 at.% of Zn for $LaFeAsO_{0.85}$ in our study.[19] The early results seem to contradict each other. It is highly possible that the Zn substitution suffered somewhat from the high volatility of Zn, resulting in an overestimation of the net Zn content.[13,20] Our recent studies in fact showed that more than 2 at.% of Zn is hardly doped into Ba(Fe,Co)$_2$As$_2$ under a regular synthesis condition.[21]

Recently, we succeeded in doping much amount of Zn into a crystal of BaFe$_{1.89}$Co$_{0.11}$As$_2$ ($T_c$ =25 K) by using a method of high-pressure and high-temperature heating. Magnetic and electrical properties of the Zn doped crystal indicate a notable $T_c$ decrease in proportional to the Zn content. Because early studies showed smaller or little $T_c$ decreases (except Ref. 19), the $T_c$ decrease is remarkable. It is hence significant to investigate the role of Zn in the crystals of BaFe$_{1.89-2x}$Zn$_{2x}$Co$_{0.11}$As$_2$ ($0 \leq x \leq 0.08$).

The nominal composition of the crystals was BaFe$_{1.87-2x}$Zn$_{2x}$Co$_{0.13}$As$_2$ ($x$=0-0.07); mixtures of BaAs (prepared as in Ref. 19), FeAs (Ref. 19), Fe (99.9%, Rare Metallic Co.), Co ($\geq$99.5%, RM), and Zn (99.99%, RM) were placed each in a boron-nitride cell, which was installed in a Ta capsule. The loaded capsule was treated at 3 GPa in a belt-type pressure apparatus at 1300 °C for 2 hrs, and the temperature was slowly decreased to 1100 °C for 2 hrs. The capsule was quenched to room temperature, followed by releasing the pressure. The as-prepared samples were kept in vacuum for 3-5 days, resulting in self-isolation of thin crystals (~0.3×0.2×0.1 mm$^3$ or smaller).

The crystal structure was investigated by powder x-ray diffraction (XRD). The tetragonal ThCr$_2$Si$_2$-type structure was found to form over the compositions from $x$ = 0 to 0.08 without traces of impurities.[22] The lattice constants were estimated from analysis of the XRD patterns (Table 1); a nearly isotropic expansion of both *a* and *c* is found, reflecting difference between the Zn-As and the Fe-As bonds as discussed in Ref. 18. In addition, a magnetic effect is possibly included in the *c*-axis expansion.[23] In addition, a shining surface of the plate-like crystal (~0.5 mm in length) was studied by XRD (Fig. 1). An orientation toward [0 0 2*n*] (*n* is integer) is obvious, indicating that the *c*-axis is perpendicular to the crystal plane. The Zn substitution was again confirmed in an electron probe micro-analysis (EPMA, JXA-8500F, JEOL) conducted on the surface (Table 1). The Co content is almost constant at ~0.11 over the compositions, while the Zn content monotonically increases with increasing $x$. Hereafter, the crystals are labeled as $BaFe_{1.89-2x}Zn_{2x}Co_{0.11}As_2$ with $x$ = 0, 0.01, 0.02, 0.03, 0.04, 0.05, 0.07 and 0.08. Note that the high-pressure method was probably essential to overcome the difficulties regarding the Zn doping.

We attempted to measure the magnetic susceptibility ($\chi$) of an individual crystal; however accurate measurements were hardly achieved. Thus, we loosely gathered crystals into a sample holder (~30 mg each composition) in Magnetic Properties Measurement System, Quantum Design for an alternative measurement. Fig. 2 shows temperature dependence of $\chi$ in a magnetic field ($H$) of 10 Oe for the crystals of $BaFe_{1.89-2x}Zn_{2x}Co_{0.11}As_2$ ($x$ = 0-0.08). The host crystal ($x$ = 0) was confirmed to show SC transition at 25 K as reported.[22,25] With increasing the Zn content, $T_c$ monotonically decreases, and the SC disappears at $x$ = 0.08 (>2 K) (Table 1).

Each crystal was carefully cleaved to a thickness of approximately 20-100 μm along the $c$-axis, and the $ab$-plane electrical resistivity ($\rho$) was measured by a standard four-point method in Physical Properties Measurement System, Quantum Design. Fig. 3 shows $T$ vs. $\rho$ for $BaFe_{1.89-2x}Zn_{2x}Co_{0.11}As_2$ ($x$ = 0-0.08); $T_c$ by $\rho$ goes down with increasing the Zn content as much as $T_c$ by $\chi$ (Table 1). This supports that the doped Zn is evenly distributed into the crystal since $T_c$ by $\rho$ is rather sensitive to the surface matter. Note that $T_c$ by $\rho$ was defined by a peak-top position of the $d\rho/dT$ curve (not shown). Besides, we define the residual resistivity $\rho_0$ by extrapolation of the $T$-linear part to zero temperature (the upturn region is excluded). Because the upturn of the resistivity curve in the highly Zn-doped crystals indicates the occurrence of localization, we tested several definitions of $\rho_0$ to avoid influence of the upturn on the $\rho_0$ estimation; however $\rho_0$ remained essentially large in any cases. The $\rho_0$ (Table 1) gradually increases with increasing the Zn content at a rate of ~76 μΩcm/%. Such a large $\rho_0$ indicates that the impurity potential by Zn is much strong, as predicted by the first principle calculation.[23] Note that the theoretical residual resistivity per 1% impurity with delta-functional strong potential is just ~20 μΩcm.[8,9] This suggests that the impurity scattering cross section is enlarged by the many-body effect.[26]

The Hall coefficient ($R_H$) at 150 K of the selected $BaFe_{1.89-2x}Zn_{2x}Co_{0.11}As_2$ crystals ($x$ = 0,

0.02, 0.04, and 0.07) was measured in the same apparatus, where $H$ was applied parallel to the $c$-axis. The data for the Zn-free crystal accesses to the early data (Table 1).[27] The $R_H$ changes little over the Zn substitution, reflecting the isoelectronic substitution of Zn for Fe. Thus, the net carrier density change is unlikely responsible for the $T_c$ decrease.

Since the impurity potential of Zn is much strong, the Zn impurity works as the unitary scattering potential that is comparable to the bandwidth. According to Ref. 8, the reduction in $T_c$ due to strong impurity ($I > 1$ eV) in the $s_\pm$ wave state is $\sim 50z$ K/%, where $z$ is the renormalization factor ($= m/m^*$; $m$ and $m^*$ are the band-mass and the effective mass, respectively). Since $m^*$ was estimated approximately between $2m$ and $3m$ by ARPES for the 122 superconductor,[28] we obtain 25 K/% (17 K/%) for $z = 0.5$ ($z = 0.33$). However, the rate for $BaFe_{1.89-2x}Zn_{2x}Co_{0.11}As_2$ is much small: 3.63 K/% is estimated by a linear fitting to the $T_c$ (by $\rho$) vs. $x$. The result quantitatively contradicts the expectation from the $s_\pm$ wave model. Meanwhile, the $s_{++}$ wave model better accounts for the result; $T_c$ is weakly suppressed by impurities due to (i) suppression of the orbital fluctuations and (ii) the strong localization effect in which the mean-free-path is comparable to the lattice spacing.[8] The (i) is because of the violation of the orbital degeneracy near the impurities and is a possible origin of the $s_{++}$ wave state.[8]

To further study the $T_c$ suppression, it is significant to calculate the pair-breaking rate $\alpha$ $=0.88z\Delta\rho_0/T_{c0}$ ($z\hbar\gamma/2\pi k_B T_{c0}$), where $\gamma$ is the electron scattering rate and $T_{c0}$ is the $T_c$ of the Zn-free crystal. On basis of the five-orbital model for the 122 system, a relation between $\gamma$ and $\Delta\rho_0$ was proposed as $\Delta\rho_0$ ($\mu\Omega$cm) $= 0.18\gamma$ (K), where $\Delta\rho_0$ is the gap between $\rho_0$ with and without Zn. In this study, we estimated $\alpha$ using $z = 0.33$ and $0.50$ ($\equiv \alpha_1$) as depicted in Fig. 4. To obtain the elastic scattering rate, we also calculated $\alpha$ by deriving $\gamma = ne^2\Delta\rho_0/2m$ ($\equiv \alpha_2$), where $n$ is the carrier number estimated from the Hall data. Both $\alpha_1$ and $\alpha_2$ data change in roughly linear; $\alpha$ is thereby estimated to be 7.64, 11.49, and 6.76 for $\alpha_1$ ($z = 0.33$), $\alpha_1$ ($z = 0.05$), and $\alpha_2$, respectively. For the $s_\pm$-wave state, the SC is expected to vanish in the range $\alpha > 0.22$ ($\alpha^\pm_c$),[8] which is remarkably much lower than the experimental values. In addition, using the relation $\alpha_3 = \hbar\Delta\rho_0/4\pi T_c\mu_0\lambda_0^2$, we obtain $\alpha_3 = 2.58$ for $\lambda_0 = 195$ nm,[5] being still very far from $\alpha^{\pm}_c$. Obviously, any pair-breaking parameter for the present superconductor is too far from $\alpha^{\pm}_c$ to support the $s_{\pm}$-wave model, indicating realization of the $s_{++}$ wave state.

It is possible that $\alpha_1$, $\alpha_2$, and $\alpha_3$ are slightly overestimated if $\Delta\rho_0$ is overestimated due to influences from such as the grain boundaries and undetected factors. For further clarification, we make additional estimation using the critical impurity concentration for the $s_{\pm}$-wave state ($n^{\pm}_{imp}$). According to the discussion in Ref. 8, Zn ($I > 1$ eV) corresponds to $n^{\pm}_{imp} \sim 0.5z/T_c$ (K). Thus, we predict $n^{\pm}_{imp}$ to be 0.01 (0.015) for $z = 0.5$ (0.33); however the experimentally determined $n_{imp}$ of 0.08 ($T_c = 0$) is much higher than the theoretical values. Thus, the discussion for $n^{\pm}_{imp}$ does not support the $s_{\pm}$–wave model for BaFe$_{1.89}$Co$_{0.11}$As$_2$ either.

The pair-breaking parameters for the α-particle irradiated NdFeAs(O,F)[29] and the proton irradiated Ba(Fe,Co)$_2$As$_2$[30] are larger than $10\alpha^{\pm}_c$ (=2.2) and $17\alpha^{\pm}_c$ (=3.8), respectively, implying that the $s_{\pm}$-wave model is unlikely for the superconductors. Recent NMR studies on the P-doped BaFe$_2$As$_2$[31] and theoretical studies on the local structure of the Fe$_2$As$_2$ layer[32] suggested a possible change of the gap symmetry depending on minute factors. Besides, a change from $d$- to $s$-wave was predicted theoretically to depend on degree of disorder.[33] To understand the pair-breaking effect comprehensively on the Fe-based superconductor, additional Zn studies over varieties of the Fe-based superconductors, including the 11, 111, 122, and 1111 systems, would be helpful from p-doped to n-doped.

In summary, we studied the Zn doping effect on the $T_c$ optimized superconductor BaFe$_{1.89}$Co$_{0.11}$As$_2$ ($T_c = 25$ K). The highest Zn level of 8 at.% was achieved by a high-pressure and high-temperature technique, resulting in a complete suppression of SC, which is remarkable. The surface Zn content by EPMA truly reflects the bulk Zn content because (i) the SC transition in the χ measurements is relatively sharp as much as that for the non-Zn doped crystal, (ii) $T_c$ by χ and $T_c$ by ρ are almost comparable over the Zn content, and (iii) the XRD lattice parameters systematically change as a bulk nature. The $T_c$ suppression rate (3.63 K/%) is, however, too low to support the $s_{\pm}$-wave model. In contrast, the $s_{++}$-wave model may better account for the result.[8,9,29,30] We should note here that early Zn studies by others coincidentally reached the same conclusion because of little $T_c$ decrease by Zn. However, this was rather likely due to overestimation of the net Zn content of the regularly synthesized polycrystalline samples.

We thank Drs. M. Miyakawa, K. Kosuda and M. Sato for valuable discussions. This research was supported in part by the World Premier International Research Center from MEXT; the Grants-in-Aid for Scientific Research (22246083) from JSPS; and the Funding Program for World-Leading Innovative R&D on Science and Technology (FIRST Program) from JSPS.

Table 1 The net Zn and Co contents, lattice parameters, $T_c$, residual resistivity, and Hall coefficient of crystals of BaFe$_{1.89-2x}$Zn$_{2x}$Co$_{0.11}$As$_2$ ($x$ = 0-0.08).

| $x$ | $x$ by EPMA | Co/Ba by EPMA | $a$ (Å) | $c$ (Å) | $T_c$ (K) by $\chi$ | $T_c$ (K) by $\rho$ | $\rho_0$ (mΩ cm) | $R_H$ (m$^3$/C) at 150 K |
|---|---|---|---|---|---|---|---|---|
| 0 | 0 | 0.113(1) | 3.955(2) | 12.976(9) | 25.0 | 25.26 | 0.26 | -2.92×10$^{-9}$ |
| 0.01 | 0.008(2) | 0.114(2) | 3.957(3) | 12.980(11) | 20.0 | 19.31 | 0.42 | |
| 0.02 | 0.021(2) | 0.109(1) | 3.963(3) | 12.983(9) | 18.5 | 18.33 | 0.40 | -2.82×10$^{-9}$ |
| 0.03 | 0.033(1) | 0.105(3) | 3.967(1) | 12.989(4) | 17.0 | 15.48 | 0.48 | |
| 0.04 | 0.044(1) | 0.112(1) | 3.968(2) | 13.002(6) | 11.0 | 11.46 | 0.57 | -4.57×10$^{-9}$ |
| 0.05 | 0.052(5) | 0.117(4) | 3.968(1) | 13.001(4) | 8.0 | 9.82 | 0.59 | |
| 0.07 | 0.073(4) | 0.106(5) | 3.972(2) | 13.026(7) | 5.5 | 7.86 | 0.76 | -4.81×10$^{-9}$ |
| 0.08 | 0.082(6) | 0.107(8) | 3.977(4) | 13.033(12) | <2 | <2 | 1.02 | |

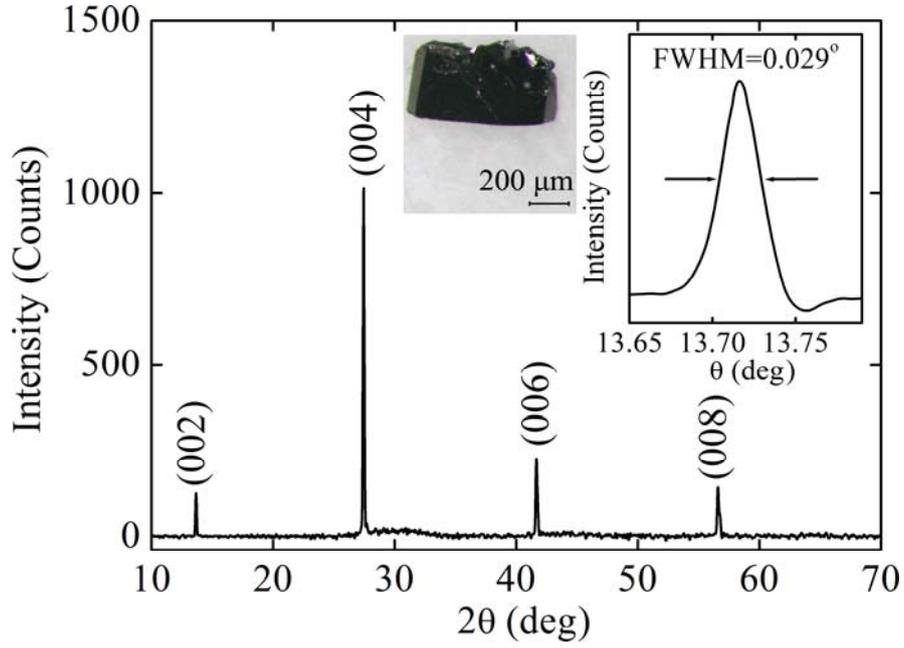

Fig. 1　XRD pattern of a crystal of BaFe$_{1.81}$Zn$_{0.08}$Co$_{0.11}$As$_2$ (EPMA).　Insets are a photograph of the crystal and rocking curve of the (004) peak.

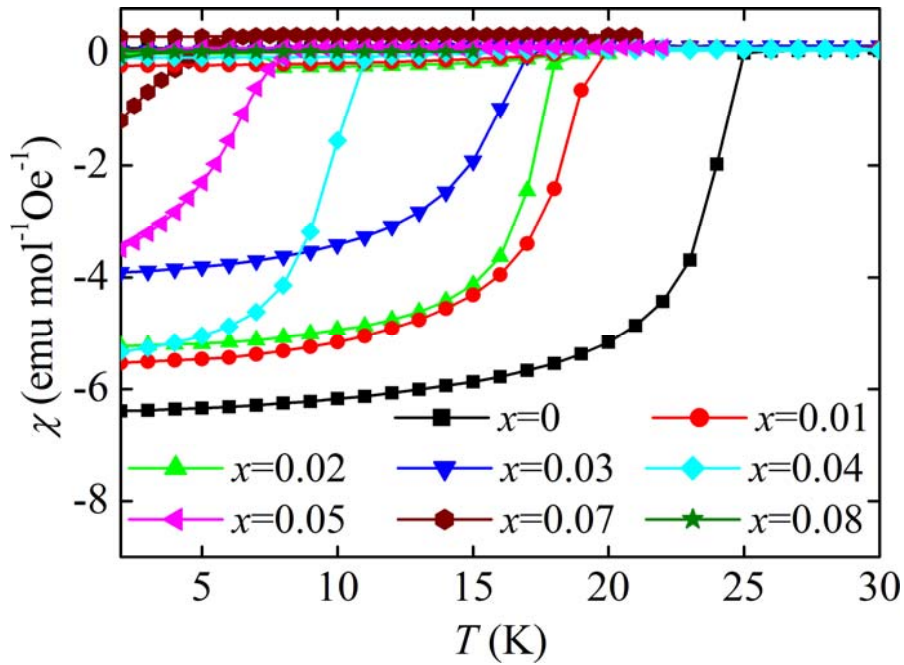

Fig. 2　χ vs. $T$ for BaFe$_{1.89-2x}$Zn$_{2x}$Co$_{0.11}$As$_2$ ($x$ = 0-0.08) at $H$ = 10 Oe.

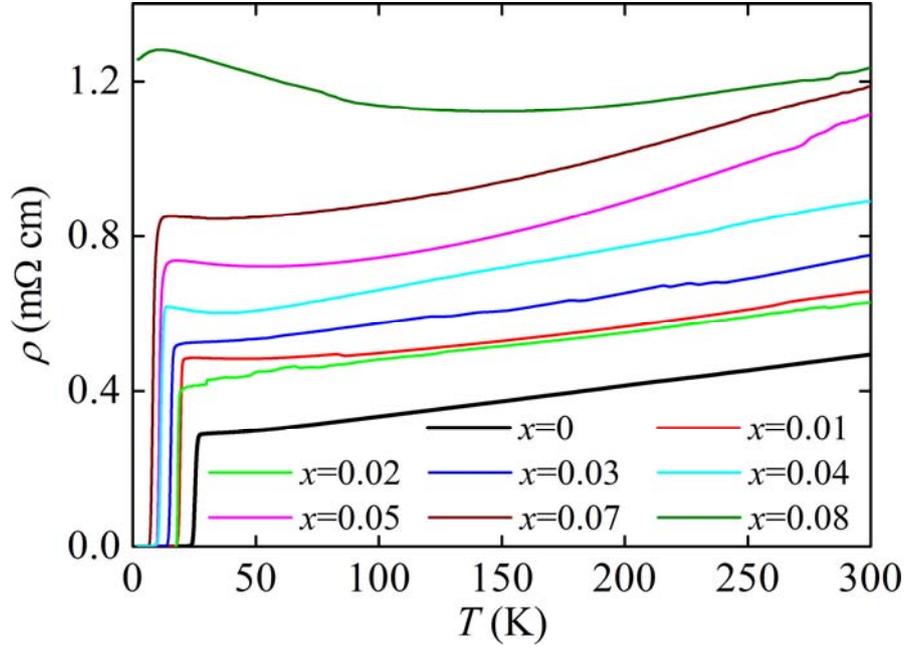

Fig. 3　$ab$-plane $\rho$ vs. $T$ for the BaFe$_{1.89-2x}$Zn$_{2x}$Co$_{0.11}$As$_2$ ($x$ = 0-0.08).

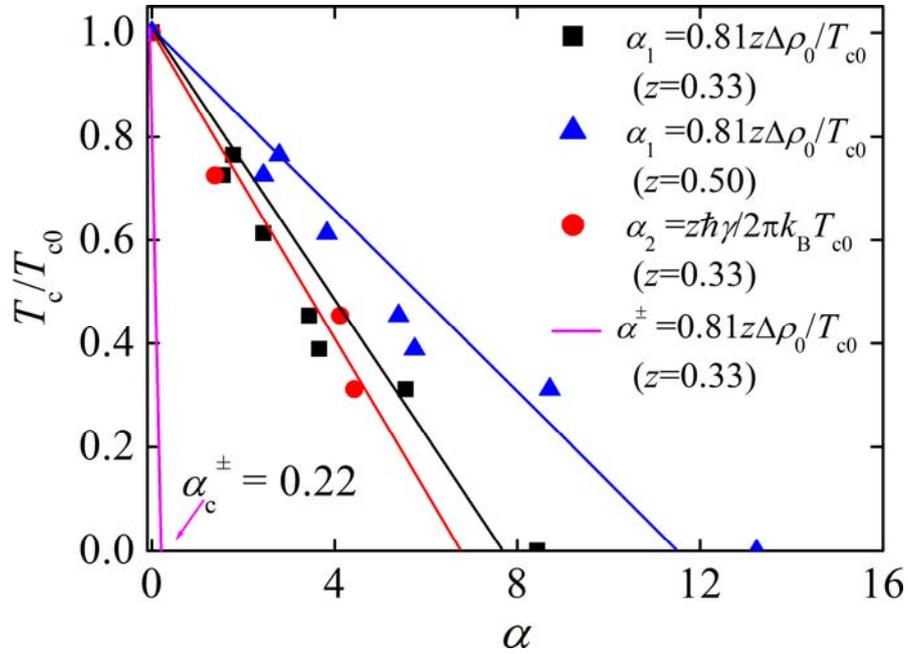

Fig. 4　$T_c/T_{c0}$ vs. $\alpha$ with various calculations for BaFe$_{1.89-2x}$Zn$_{2x}$Co$_{0.11}$As$_2$ ($x$ = 0-0.08).